\documentclass[12pt]{article}
\usepackage{graphicx}
\usepackage{hyperref}
\usepackage{subfigure}
\usepackage[utf8]{inputenc}
\usepackage[square,sort,comma,numbers]{natbib}

\usepackage{geometry}
\newcommand\pubnumber{}
\newcommand\pubdate{\today}

\def\institute{School of Physics and Astronomy\\
The University of Glasgow, Glasgow, UK}
\def\Title#1{\begin{center} {\Large #1 } \end{center}}
\def\Author#1{\begin{center}{ \sc #1} \end{center}}
\def\Address#1{\begin{center}{ \it #1} \end{center}}

\newcommand\pubblock{\rightline{\begin{tabular}{l} \pubnumber\\
         \pubdate  \end{tabular}}}
\newenvironment{Abstract}{\begin{quotation}  }{\end{quotation}}
\newenvironment{Presented}{\begin{quotation} \begin{center} 
             PRESENTED AT\end{center}\bigskip 
      \begin{center}\begin{large}}{\end{large}\end{center} \end{quotation}}
\def\Acknowledgements{\bigskip  \bigskip \begin{center} \begin{large}
             \bf ACKNOWLEDGEMENTS \end{large}\end{center}}

\newcommand{\ttbar}{\ensuremath{t\bar{t}}}
\newcommand{\pt}{$p_{\mathrm{T}}$}
             
\begin{document}
\begin{titlepage}
\pubblock

\vfill
\Title{Top2018: Experimental Summary}
\vfill
\Author{ Mark Owen}
\Address{\institute}
\vfill
\begin{Abstract}
Top quark physics continues to be an exciting and fast moving research area. The large statistics provided by the LHC
are allowing us to measure processes never observed before and to develop new methods to improve the precision
for the ``bread-and-butter" measurements. Summarising more than thirty talks in a concise way is something of a challenge
and hence this document is my own personal biased selection of the many interesting results that were discussed at the workshop.
\end{Abstract}
\vfill
\begin{Presented}
$11^\mathrm{th}$ International Workshop on Top Quark Physics\\
Bad Neuenahr, Germany, September 16--21, 2018
\end{Presented}
\vfill
\end{titlepage}
\def\thefootnote{\fnsymbol{footnote}}
\setcounter{footnote}{0}

\section{Introduction}

Experimental measurements of the top-quark have, by now, a substantial history, dating back to the discovery in 1995~\cite{Abe:1995hr,D0:1995jca}.
The Large Hadron Collider is now delivering on one of its goals by acting as a top-quark factory and producing huge datasets for analysis
by the experiments. The Top series of workshops has for the past years been the venue for the presentation and discussion of new results
and this year's workshop was no different. These proceedings represent the author's personal view of the experimental results presented at the Top 2018 workshop.
The proceedings start with a discussion of top-quark pair cross-section measurements in Section~\ref{s:crosssection}. Section~\ref{s:properties} takes a look
at developments in top properties measurements, with an eye on the connection to differential cross-sections. Single-top quark production is discussed in Section~\ref{s:singletop}
and the trend to look for rare $\ttbar X$~processes is explored in Section~\ref{s:ttX}. I borrow from one of the theoretical speakers to try to highlight a few
searches with top quarks in Section~\ref{s:search} before concluding with a look to the future in Section~\ref{s:future}.

\section{Top cross-section measurements: mass production}
\label{s:crosssection}

\begin{figure}[htb]
\centering
\includegraphics[width=0.4\textwidth]{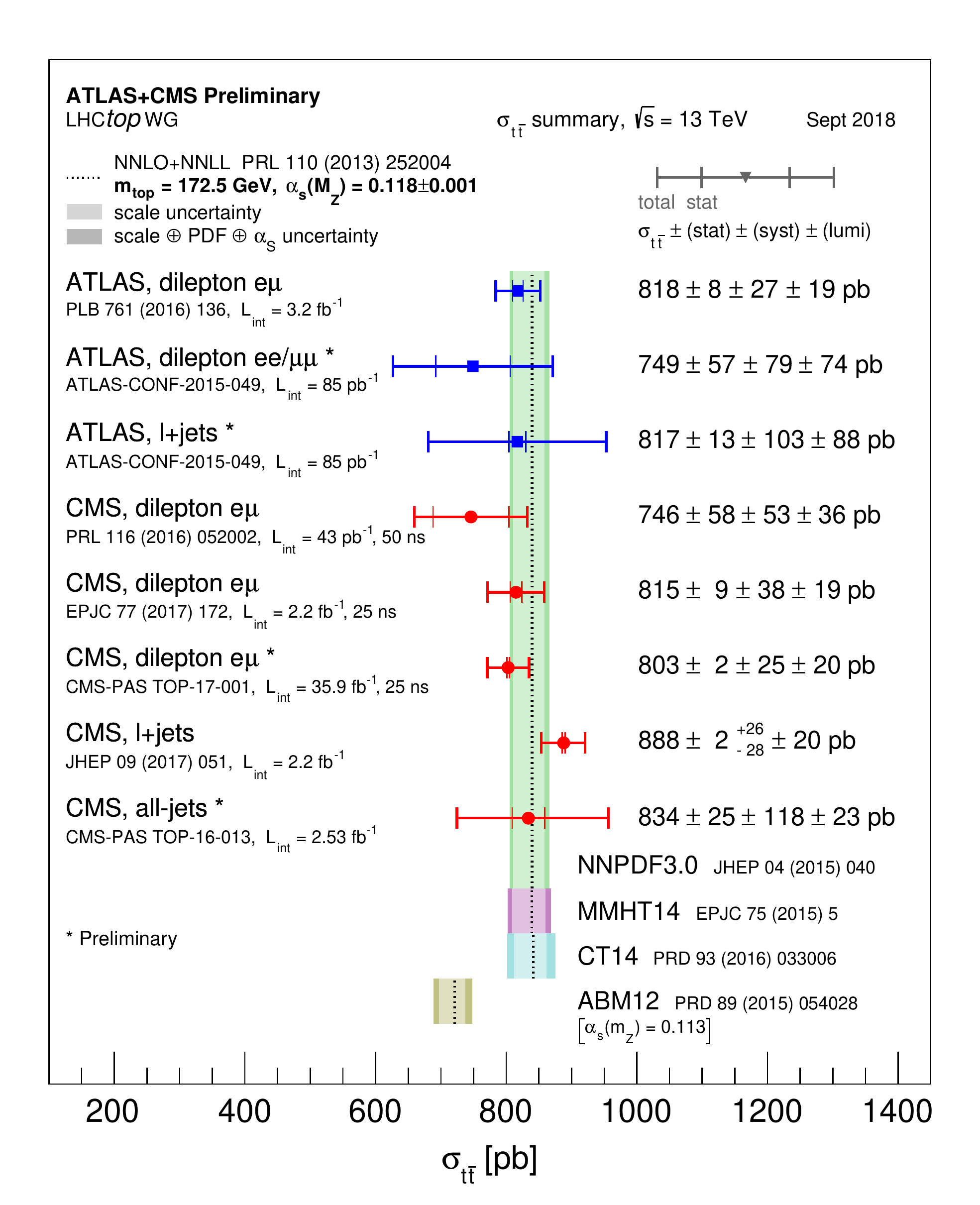}
\caption{Summary of the measurements of the inclusive $t\bar{t}$~production cross-section measurements at $\sqrt{s}=13$~TeV.}
\label{f:ttbarinclusiveXS13}
\end{figure}

The huge statistics available at the LHC has for several years meant that the measurement of the top-quark pair cross-section
is no longer simply a case of counting events to measure a single number. Instead the collaborations have developed sophisticated methods
to measure the inclusive cross-section with impressive precision and now produce multi-dimensional differential cross-section measurements
that can test our best Standard Model (SM) predictions in great detail. The state-of-the-art for inclusive cross-section measurements at $\sqrt{s}=13$~TeV
is shown in Figure~\ref{f:ttbarinclusiveXS13}. Both experiments reach a precision of $4\%$~in the dilepton channel~\cite{Aaboud:2016pbd,Sirunyan:2018goh},
however it is worth noting that the analysis techniques deployed
are rather different and (possibly consequently) the dominant systematics are also different. To me, this indicates that while the inclusive cross-section
measurements are clearly systematics limited, we can still expect that further work and new ideas have the potential to improve the precision on these benchmark measurements.
One place where systematic uncertainties are not dominant is at LHCb, who presented a new cross-section measurement of top production at $13$~TeV~\cite{Aaij:2018imy}.
This is a nice development and opens the possibility for higher precision in the years ahead in a region of phase space not fully covered by ATLAS and CMS.

\begin{figure}[htb]
\centering
\subfigure[]{
\includegraphics[width=0.38\textwidth]{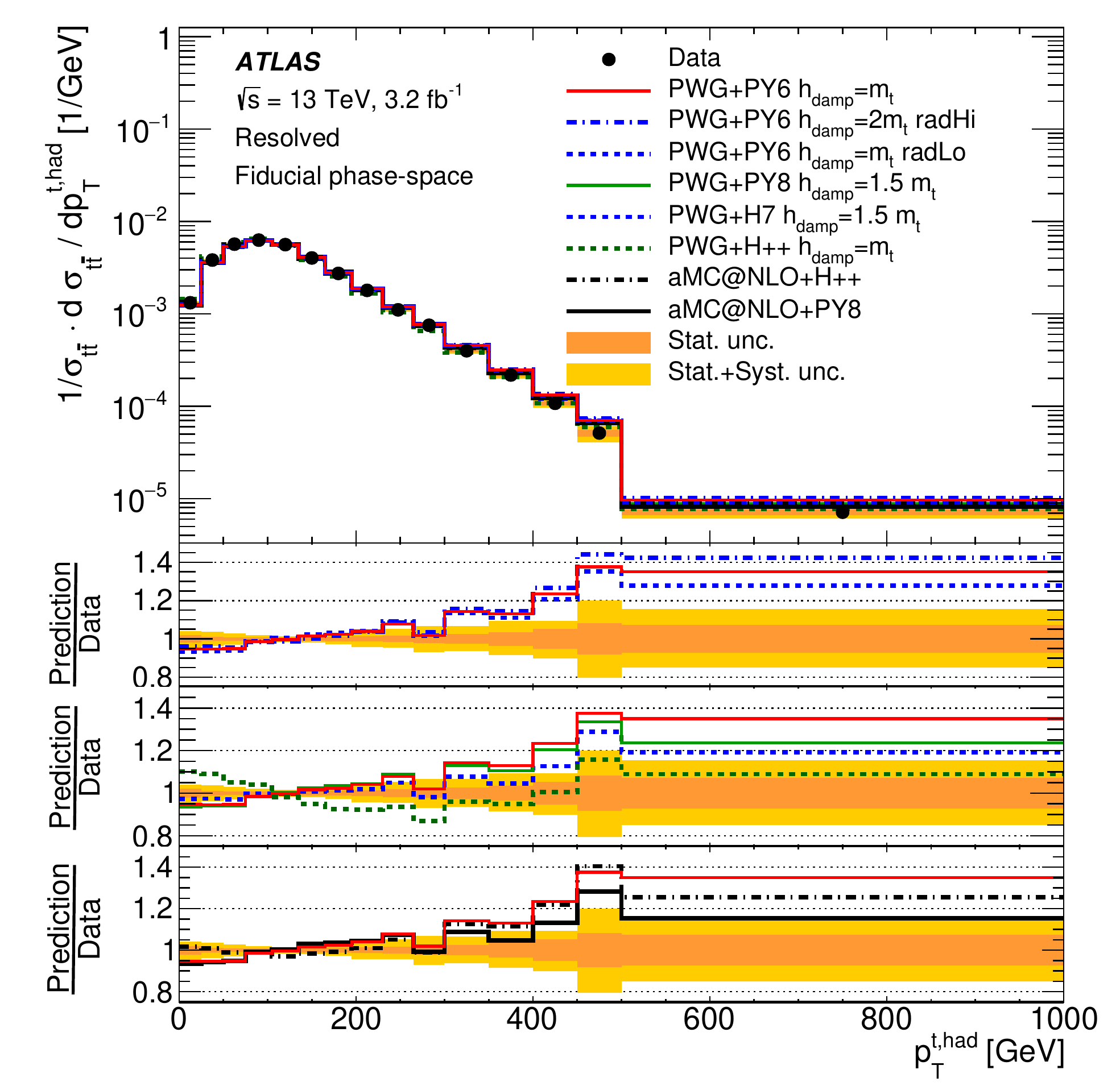}}
\subfigure[]{
\includegraphics[width=0.39\textwidth]{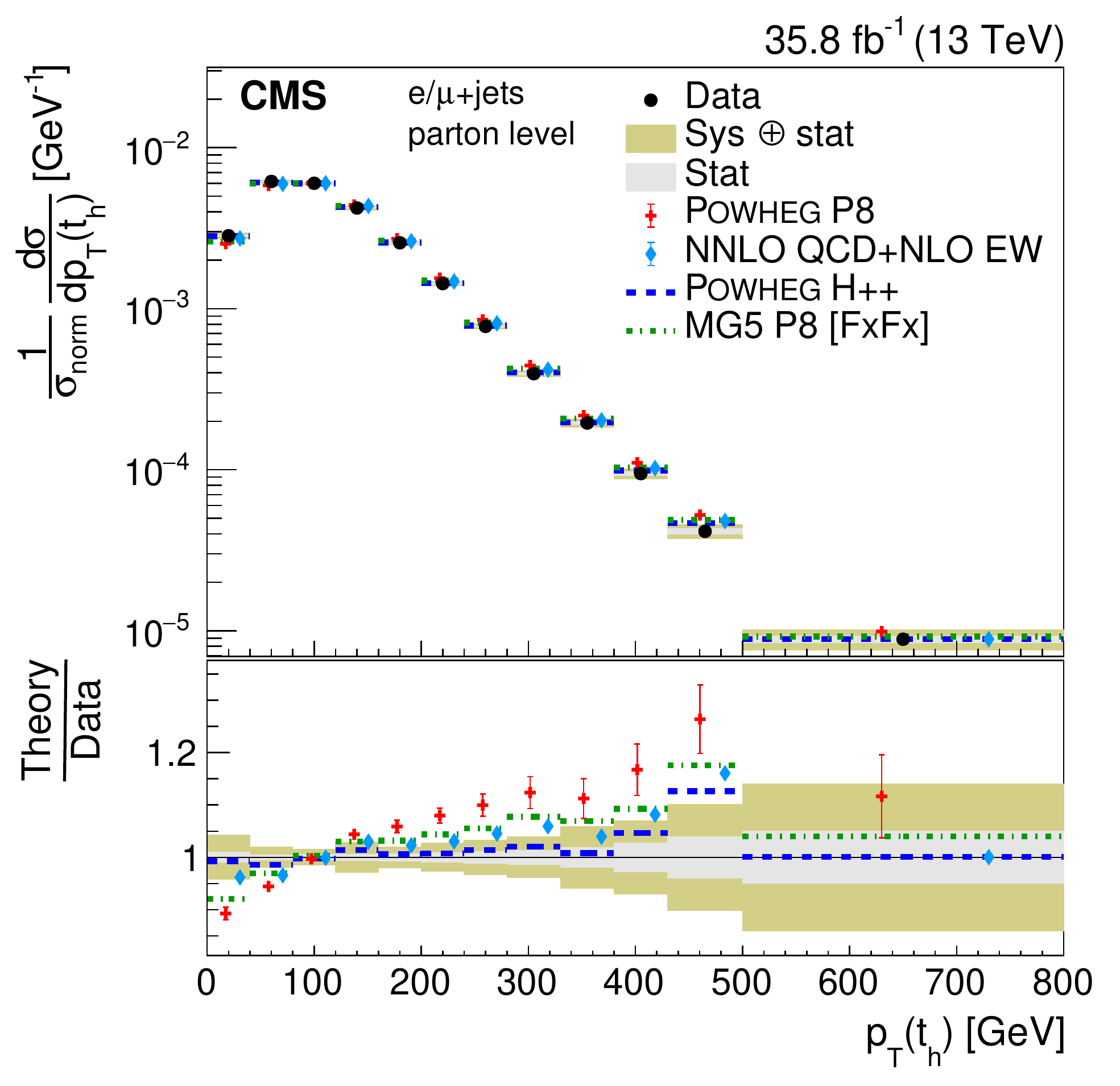}}
\caption{Measurements of the top-quark \pt~distribution from (a) ATLAS~\cite{Aaboud:2017fha} and (b) CMS~\cite{Sirunyan:2018wem}.}
\label{f:toppT}
\end{figure}

Measurements of the differential cross-section of $t\bar{t}$~production have now entered a ``mass production" phase. Indeed, in the three
excellent presentations covering the results, no fewer than $54$~unfolded differential distributions were presented. This represents
a wealth of measurements to confront with the theory. In general, reasonable agreement with Monte Carlo predictions that use NLO
matrix-elements is seen. There continues to be some disagreement in the top-quark \pt~distribution, as demonstrated in Figure~\ref{f:toppT}.
This agreement seems to be improved by the NNLO predictions, however this is not yet fully established. This perhaps illustrates one of the challenges
in the cross-section area, how to turn this very impressive array of measurements into concrete answers to physics questions such as:
does the most precise SM calculation agree with the data and do the ATLAS and CMS data agree with each other? This is something I hope
progress can be made on in the next years, along with analysing the full run-2 dataset. An important part of the cross-section programme is
to ensure that the measurements are made available for phenomenological studies. Here, it was very pleasing to see that many analyses
are now routinely made available in the Rivet format~\cite{Buckley:2010ar} and the data are published to the HepData repository~\cite{Maguire:2017ypu}.

%The single-top $t$-channel process is now joining $\ttbar$~production in the precision regime. One of the new results presented at the workshop
%by CMS showed an impressive measurement of this process, in particular reaching a precision of 

\section{From differential cross-sections to top-quark properties}
\label{s:properties}

\begin{figure}[htb]
\centering
\subfigure[]{
\includegraphics[width=0.38\textwidth]{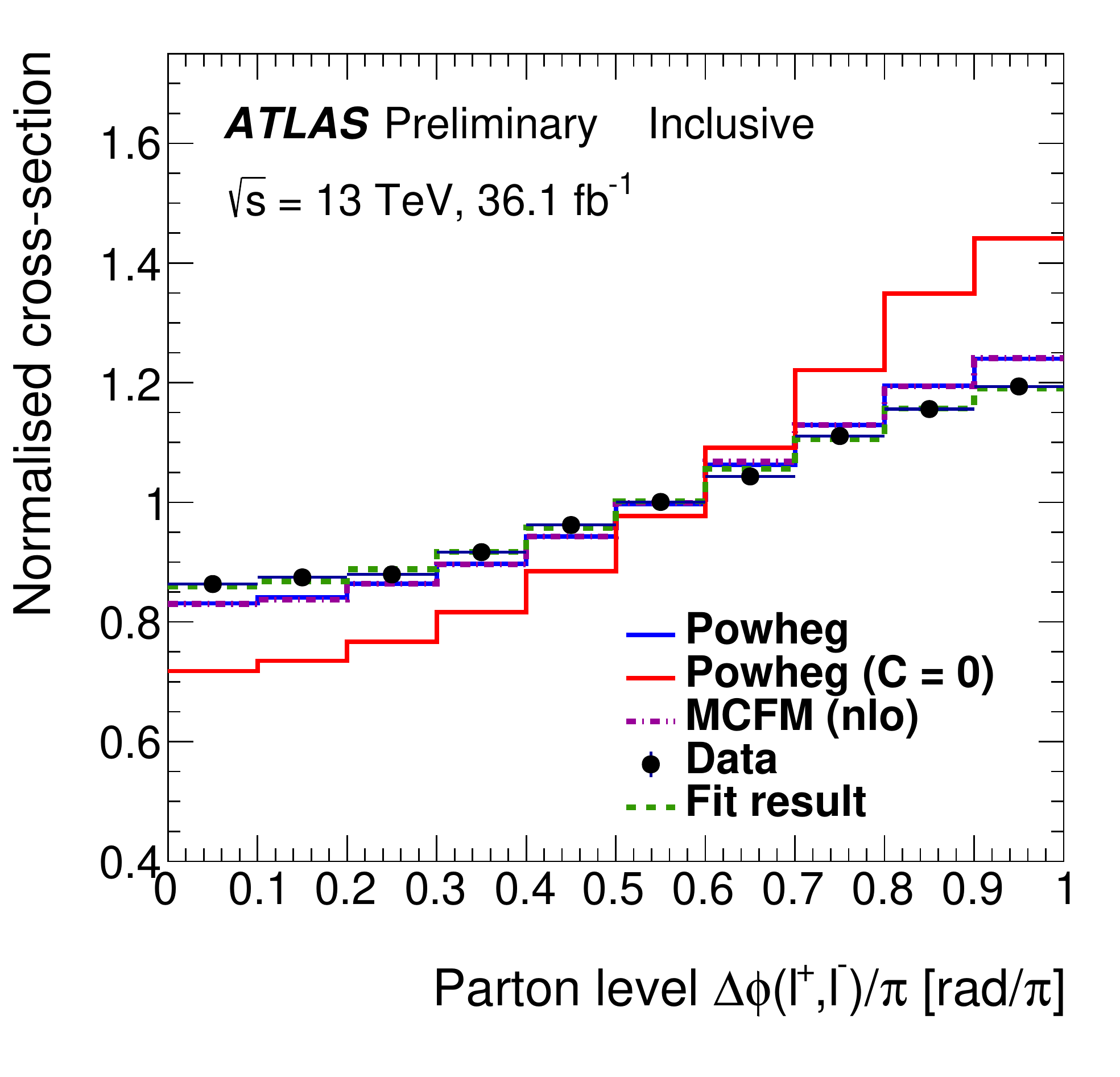}
}
\subfigure[]{
\includegraphics[width=0.37\textwidth]{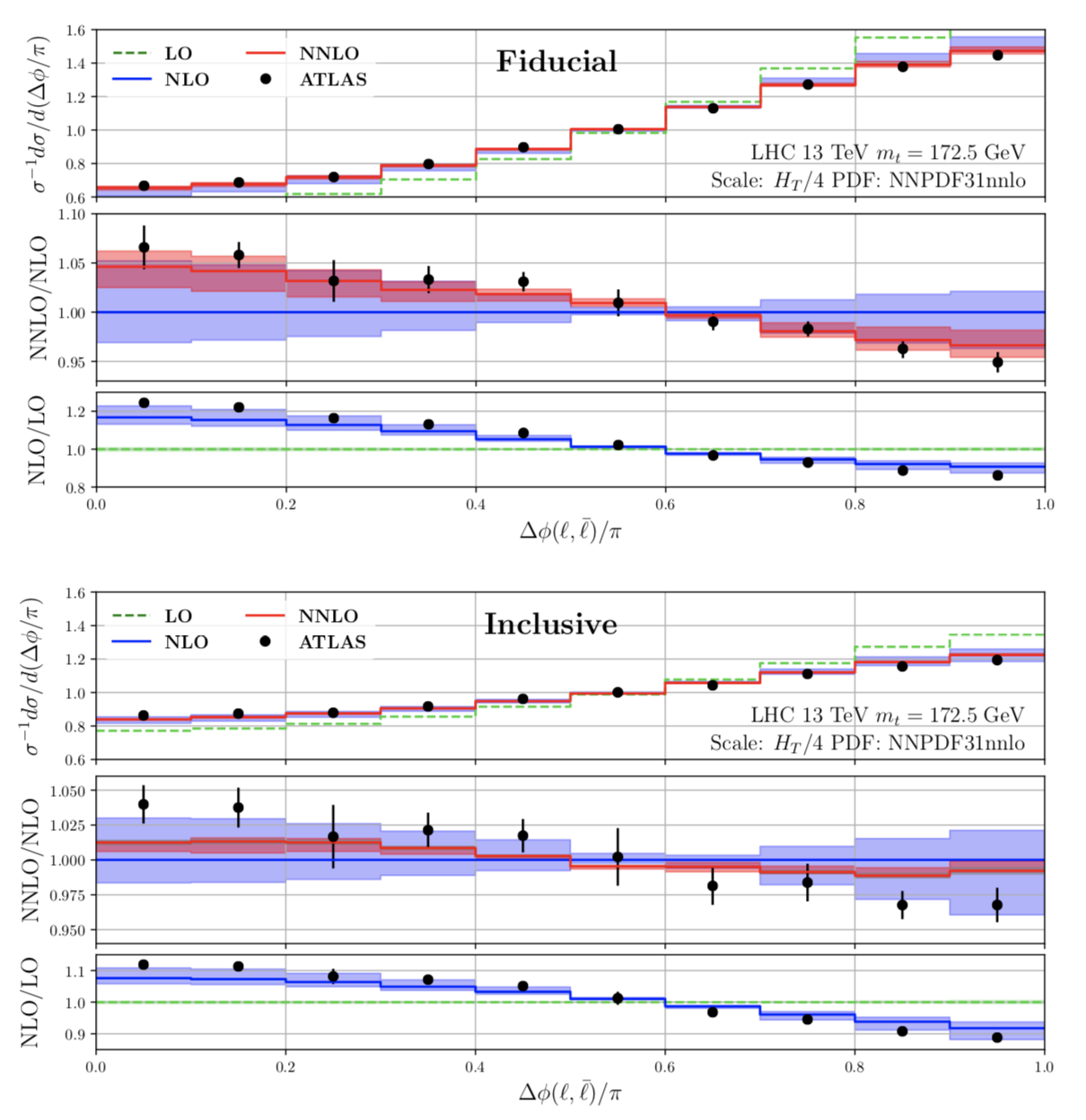}
}
\caption{(a) The measurement of the $\Delta\phi(e,\mu)$~distribution by ATLAS~\cite{ATLAS-CONF-2018-027} is compared to NLO MC predictions, including one with zero spin correlations (red).
(b) Comparisons of the ATLAS measurement to recent NNLO predictions~\cite{Behring:2019iiv}, where the upper panel shows the fiducial measurement and the lower panel shows the full
phase-space measurement.
}
\label{f:spin}
\end{figure}

The correlation between the spins of top-quarks in $pp$~collisions can be predicted from QCD calculations. Since top-quarks decay before being detected, 
this spin correlation cannot be measured directly, but must be inferred from measurements of the decay products. The recent measurement of spin
correlation from ATLAS~\cite{ATLAS-CONF-2018-027} provides an interesting example of the connection between differential cross-section measurements and top-quark properties.
In this case, ATLAS measures the opening angle in the azimuthal plane between the two leptons in dilepton top-quark pair events. This variable is
highly correlated with the spin correlation of the top-quark pair and the degree of spin correlation is then extracted from the differential cross-section
measurement. The measured cross-section is shown in Figure~\ref{f:spin}. The data clearly exclude the no-spin scenario, however they also show
notable disagreement with the NLO SM prediction, which is at the level of three standard deviations. The source of this discrepancy is as yet not
established, but could for example be due to higher order QCD effects, or perhaps more speculatively, the impact of some new physics contribution.
The benefit of providing fully unfolded differential cross-sections should be that they can be compared very quickly to improved theoretical predictions.
This seems to be realised in the case of the spin correlation measurement, with the data being compared to new NNLO differential predictions
shortly after the workshop~\cite{Behring:2019iiv}. The picture here is still not fully resolved, with some differences between the fiducial and
full phase-space comparisons visible in Figure~\ref{f:spin}. Hopefully, close collaboration between theorists and experimentalists will provide a clearer
picture in the near future.

Perhaps the most measured property of the top quark is its mass, which is a crucial parameter in terms of understanding the self-consistency of the
SM. The measurement of the mass is now tackled from multiple angles and one interesting technique that has been refined over
the last couple of years is to extract the top quark mass from cross-section measurements. The latest CMS cross-section measurement has recently been
used to do such an extraction, yielding $m_t = 173.2 ^{+2.1}_{-2.3}$~GeV~\cite{Sirunyan:2018goh} when using the NNPDF3.1 PDF set.
Unfortunately, it does look highly challenging to improve such an extraction to get below $1$~GeV uncertainty, to approach the precision of direct measurements.
An interesting extension of the extraction from the inclusive cross-section measurement was presented by ATLAS~\cite{Aaboud:2017ujq} where differential
distributions of the leptons in dilepton top-quark events produced in 8 TeV collisions are used to extract the top quark mass. The measurement yields $m_t = 173.2 \pm 1.7$~GeV.
%as shown in Figure~\ref{f:topmass}.
It is worth noting that the largest systematic uncertainty comes from the scale uncertainties in the theoretical prediction ($1.2$~GeV) and that the statistical
uncertainty is still significant at $0.9$~GeV. The predictions used
in the extraction are calculated at NLO in QCD and hence the recent availability of NNLO predictions, along with the large statistics of the run-2 dataset, offer the
possibility for a significant improvement in the uncertainty in the future.

%\begin{figure}[htb]
%\centering
%\subfigure[]{
%\includegraphics[width=0.4\textwidth]{ATLASdilepdiff_mtop_fig_21.pdf}}
%\subfigure[]{
%\includegraphics[width=0.35\textwidth]{CMS-TOP-17-008_Figure_002-a.pdf}}
%\caption{(a) Results of the top mass extracted from the ATLAS measurement of differential cross-sections in the dilepton channel. (b) Reconstructed invariant top-quark mass distribution
%in the all-hadronic channel from CMS.}
%\label{f:topmass}
%\end{figure}

The most precise top mass measurements are still obtained by reconstructing the decay products and fitting distributions that have a high sensitivity to the top mass.
The CMS collaboration is leading the way in applying these techniques to the run-2 data, presenting measurements using both the lepton-plus-jets~\cite{Sirunyan:2018gqx} and all-hadronic
channels~\cite{Sirunyan:2018mlv}.
%The reconstructed top mass distribution in the all-hadronic channel is shown in Figure~\ref{f:topmass}, showing the excellent resolution and low background achieved
%in that channel.
The run-2 measurements do not yet reach the precision obtained in run-1, indicating that there is plenty of work ahead in the collaborations to break the $0.5$~GeV barrier.
Finally, the ATLAS experiment recently produced a new lepton-plus-jets top mass measurement using the 8 TeV data, which when combined with the other ATLAS measurements yields
$m_t= 172.69 \pm 0.48$~GeV~\cite{Aaboud:2018zbu}. Along with the CMS run-1 combined measurement of $172.44 \pm 0.49$~GeV~\cite{Khachatryan:2015hba},
this suggests to me that it is highly desirable to pursue an updated top mass combination.

\section{Single top, but not alone}
\label{s:singletop}

The three leading-order single-top quark production processes, $t-$channel, $s-$channel and $tW$~production all provide direct access to the
$Vtb$~coupling. The $t-$channel process has entered a precision regime experimentally and a new measurement by CMS of the inclusive
cross-section and ratio of the top and anti-top cross-sections ($R_t$) was presented~\cite{Sirunyan:2018rlu}. The measurement uses a sophisticated fit to constrain
the systematic uncertainties. The cross-section measurements suffer from a large modelling uncertainty, while the $R_t$~measurement reaches a precision
of $2.8\%$, improving on the run-1 measurements.

\begin{figure}[htb]
\centering
\subfigure[]{
\includegraphics[width=0.35\textwidth]{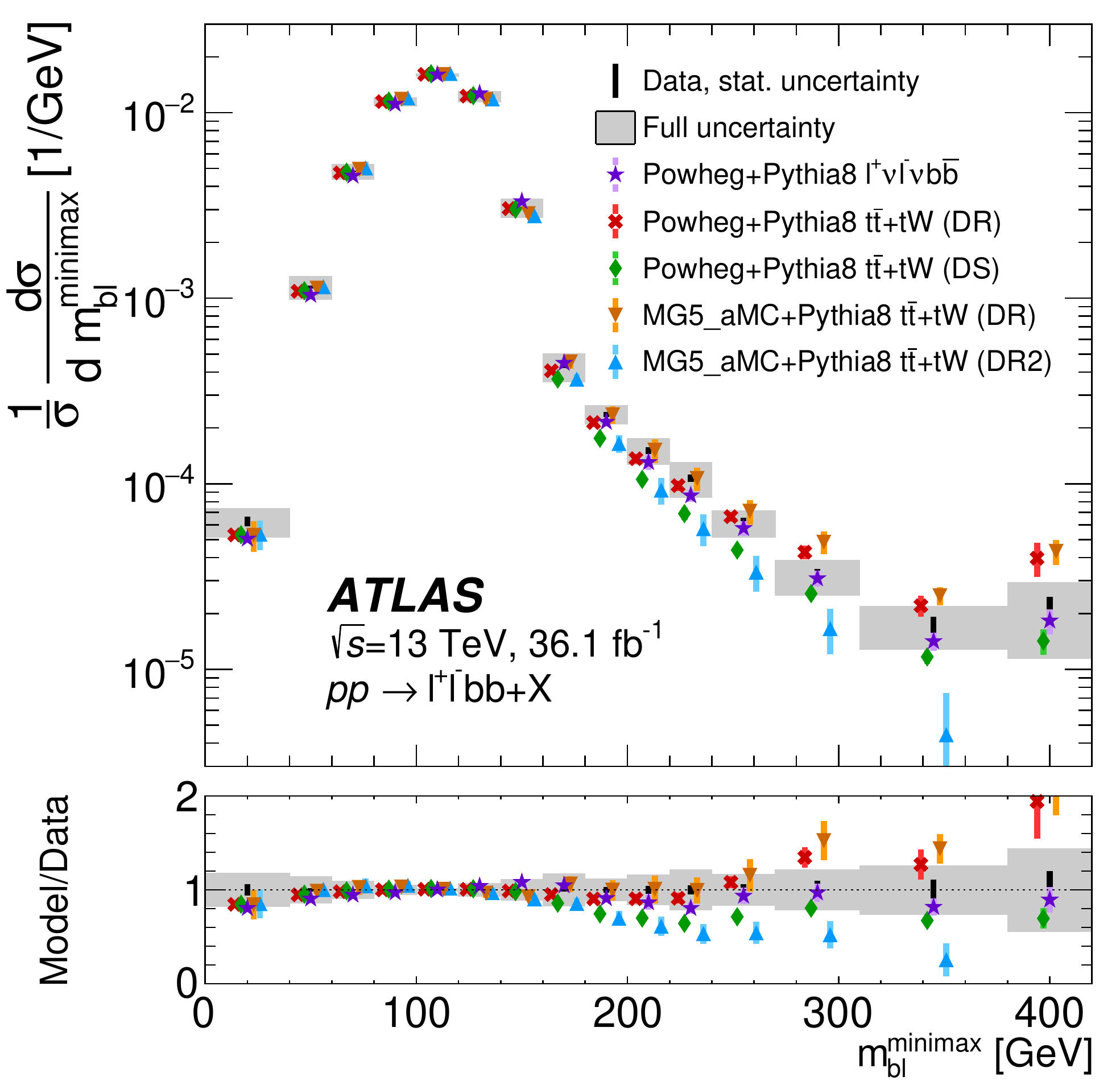}}
\subfigure[]{
\includegraphics[width=0.44\textwidth]{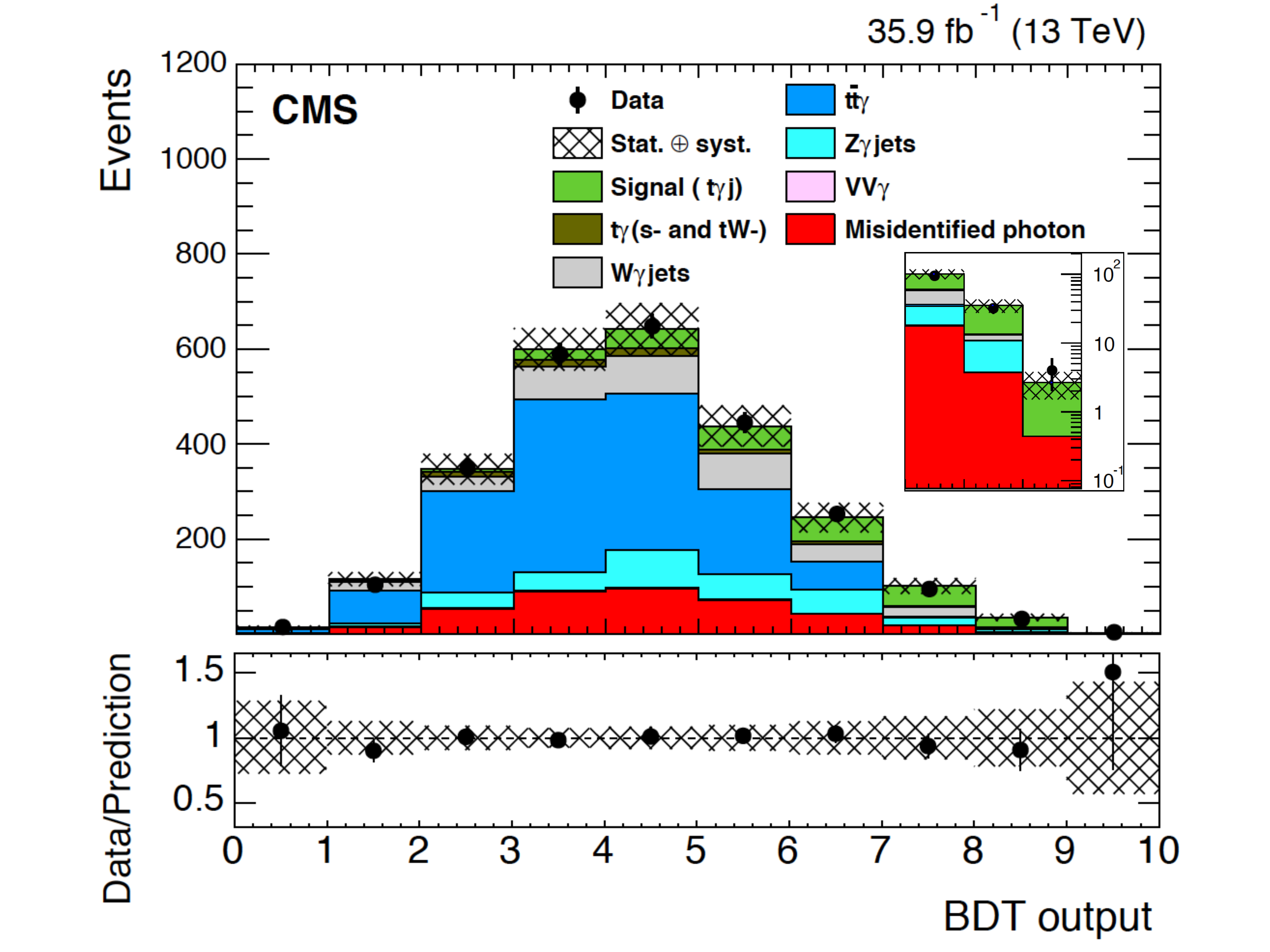}}
\caption{(a) The measurement of the $m_{b\ell}^{\mathrm{minimax}}$~distribution is compared to several theoretical predictions (see~\cite{Aaboud:2018bir} for
details). The purple points show the prediction from the $WbWb$~MC. (b) BDT output used for the measurement of single-top production in association
with a photon~\cite{Sirunyan:2018bsr}.}
\label{f:WbWb}
\end{figure}

A fascinating feature of the $tW$~process is that it cannot be disentangled from the \ttbar~process beyond leading-order. On the theoretical side, this issue has
been addressed by calculating the full $WbWb$~process, including off-shell and interference effects~\cite{Denner:2010jp,Bevilacqua:2010qb,Jezo:2016ujg}.
The ATLAS collaboration presented a measurement~\cite{Aaboud:2018bir} that is designed to be sensitive to the region where the off-shell and interference effects are relevant.
This is achieved by selecting events with two leptons and two $b$-jets and measuring the differential cross-section as a function of $m_{b\ell}^{\mathrm{minimax}}$~\cite{Aaboud:2018bir}. The measured distribution can be seen in Figure~\ref{f:WbWb} and it is observed that the best agreement with the data is seen for the full $WbWb$~calculation.

The large statistics of the LHC allow the possibility to detect single top-production in association with gauge bosons. The last Top workshop saw first evidence
for the $tZ$~process~\cite{Sirunyan:2017nbr,Aaboud:2017ylb} and this workshop saw the first measurement of single-top production in association
with a photon presented by CMS~\cite{Sirunyan:2018bsr}. The boosted decision tree (BDT) output used to extract the signal can be seen in Figure~\ref{f:WbWb},
which shows the high purity obtained at high BDT values. I think these measurements mark the start of a programme to measure in detail these associated production modes,
including in the future differential cross-section measurements.

\section{Top quark pairs + X}
\label{s:ttX}

\begin{figure}[htb]
\centering
\subfigure[]{
\includegraphics[width=0.46\textwidth]{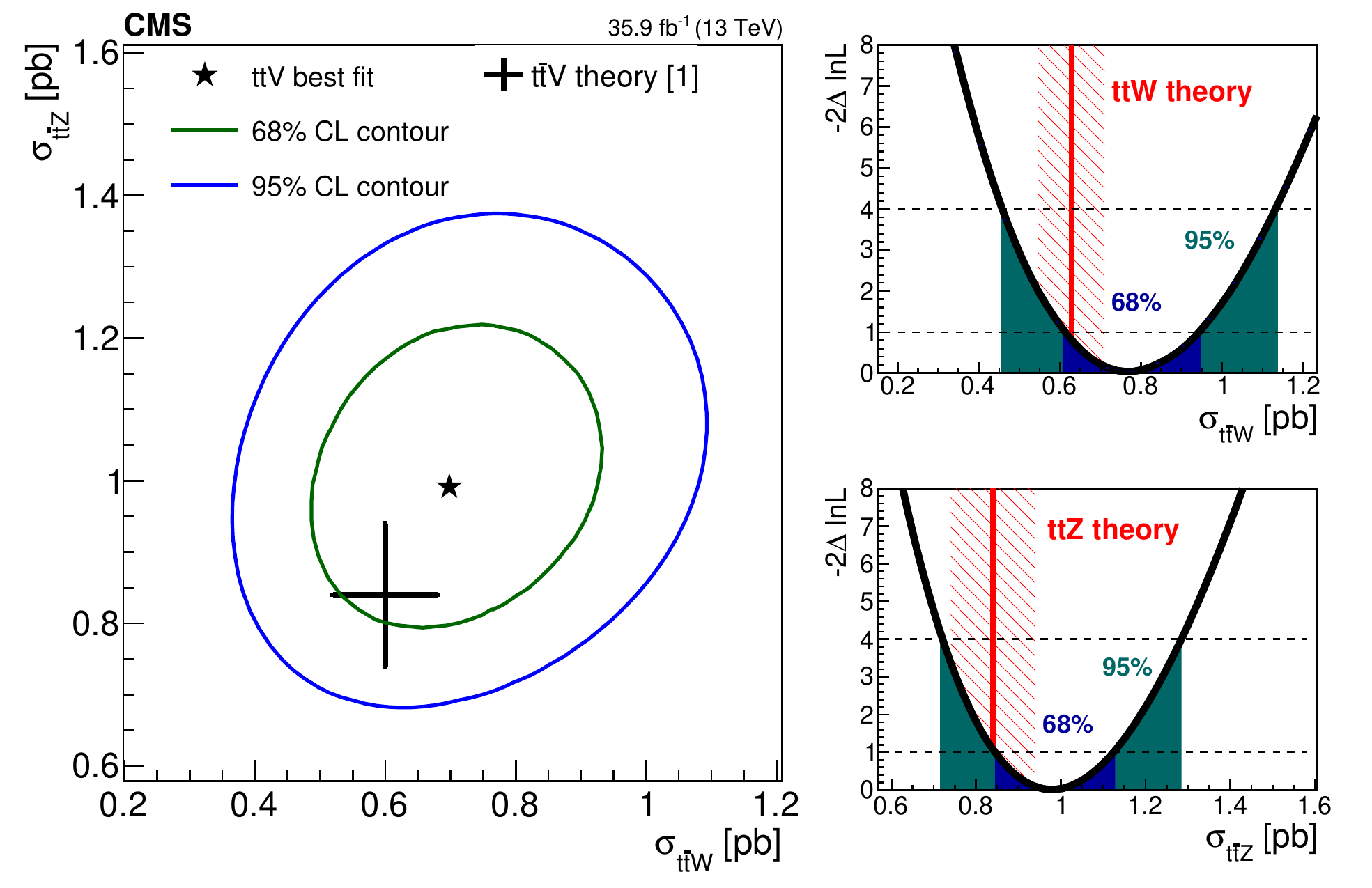}}
\subfigure[]{
\includegraphics[width=0.31\textwidth]{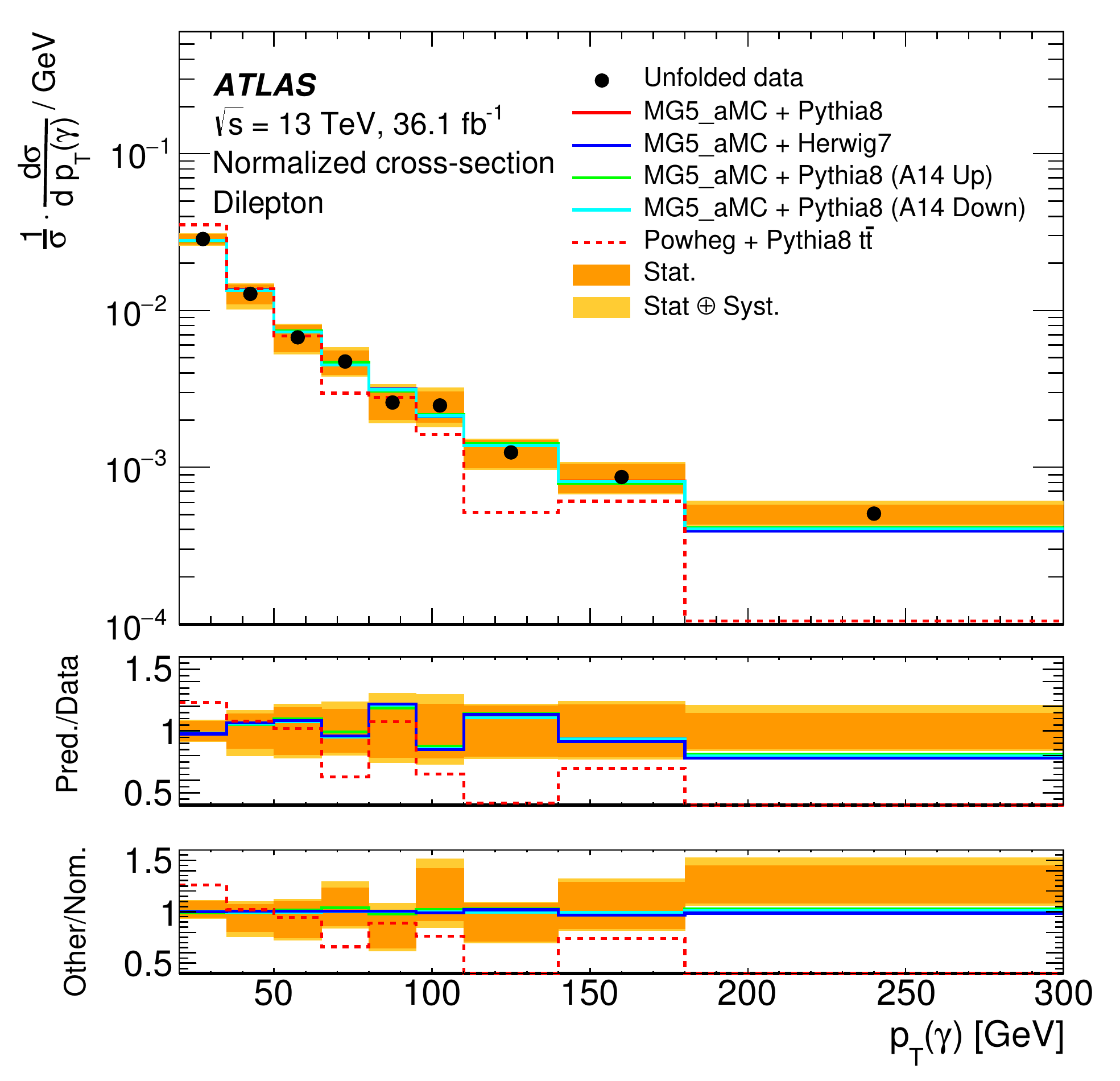}}
\caption{(a) Results of the CMS measurement of the $\ttbar W$~and $\ttbar Z$~cross-sections~\cite{Sirunyan:2017uzs} are compared with the theoretical predictions.
(b) ATLAS measurement of the differential cross-section of the $\ttbar \gamma$~process as a function of the transverse momentum of the photon~\cite{Aaboud:2018hip}.
}
\label{f:ttZ}
\end{figure}

As discussed already in Section~\ref{s:singletop}, the high statistics of the LHC allows the possibility to measure various associated production processes.
$\ttbar Z$~is perhaps one of the most interesting of such processes, since it allows direct access to the top-$Z$~boson coupling.
The latest CMS measurement of $\ttbar Z$~\cite{Sirunyan:2017uzs} reaches a precision $14\%$ and as seen in Figure~\ref{f:ttZ} is in good agreement
with the SM prediction. What is particularly striking is that the measurement now reaches
a point where the systematic uncertainty is slightly larger than the statistical uncertainty. Further significant improvements in the precision will therefore
now require reductions of the systematic uncertainties, however here we can be optimistic that the progress made in continually improving the $\ttbar$~cross-section
measurement (see Section~\ref{s:crosssection}) is a sign that similar improvements will be achievable in these rarer production modes.
The high statistics should also allow differential measurements of the associated production processes to be made and steps in this direction
were seen from ATLAS, with a new measurement of the $\ttbar \gamma$~process at 13 TeV~\cite{Aaboud:2018hip}. The paper includes 
several differential distributions and Figure~\ref{f:ttZ} shows the measurement as a function of the transverse momentum of the photon.

\begin{figure}[htb]
\centering
\subfigure[]{
\includegraphics[width=0.45\textwidth]{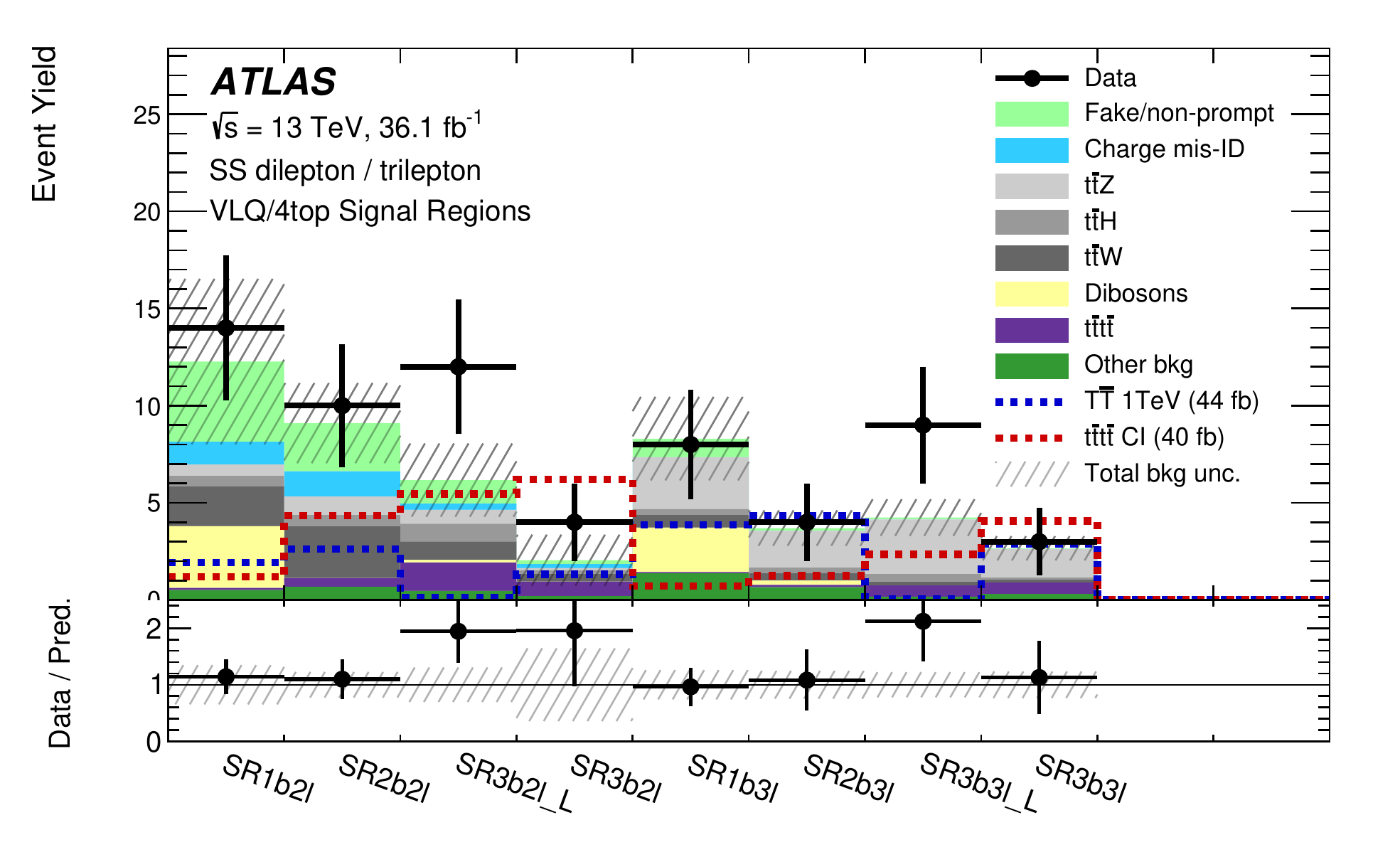}}
\subfigure[]{
\includegraphics[width=0.28\textwidth]{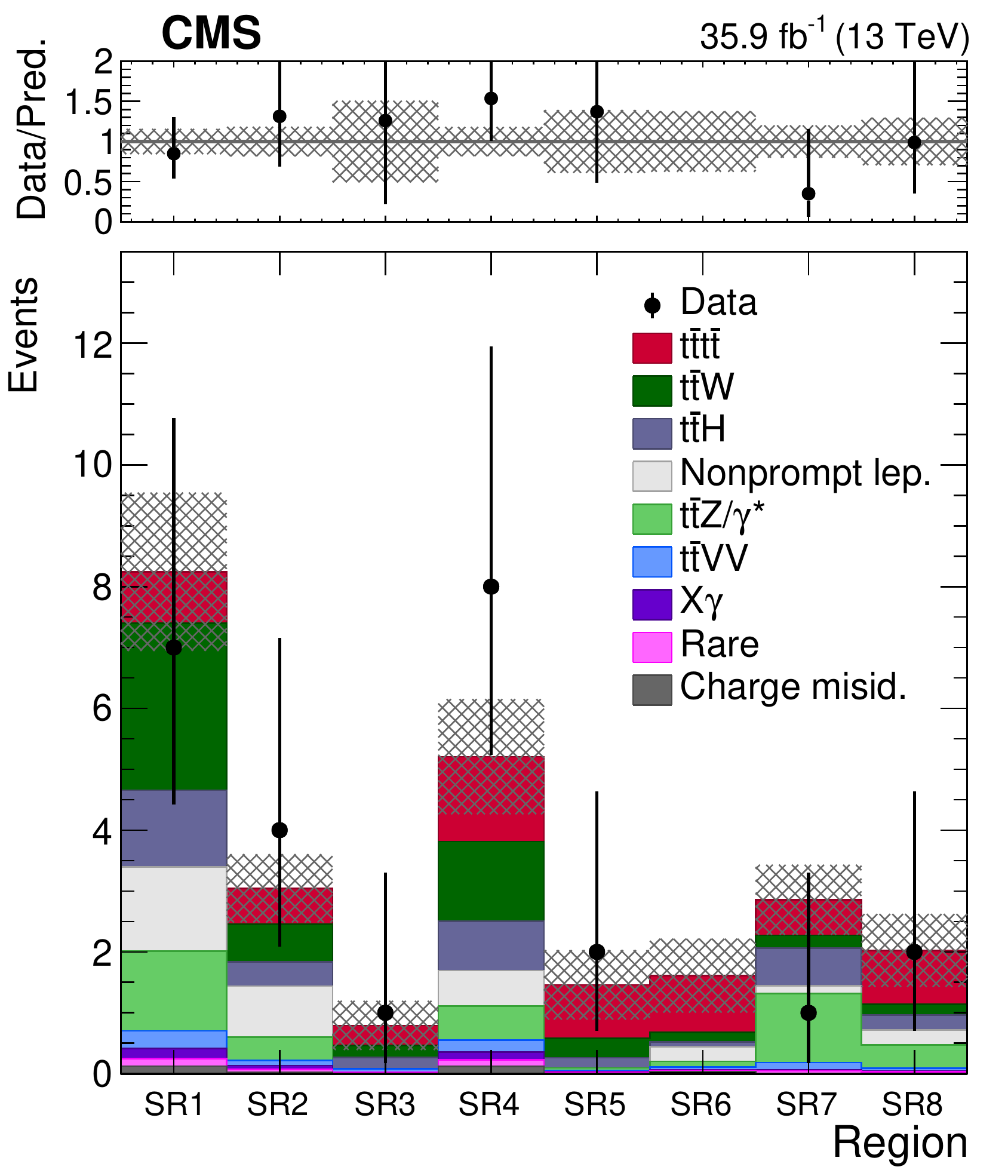}}
\caption{Distributions showing the number of events observed in the signal regions for the four-top searches at (a) ATLAS~\cite{Aaboud:2018xpj} and (b) CMS~\cite{Sirunyan:2017roi}.}
\label{f:fourtop}
\end{figure}

The measurements of these rare production processes naturally lead one to consider measuring top-pairs produced in association with more top-pairs,
leading to a four-top final state. As well as being interesting from a QCD point of view, this topology could be produced in various new physics models.
The signal regions for the latest searches~\cite{Sirunyan:2017roi,Aaboud:2018xpj} for this process are shown in Figure~\ref{f:fourtop}. As yet there is no solid evidence for the process, however
the full run-2 LHC dataset should offer the possibility to have a first look at four-top events - perhaps something to look out for at Top2019.

\begin{figure}[htb]
\centering
\subfigure[]{
\includegraphics[width=0.35\textwidth]{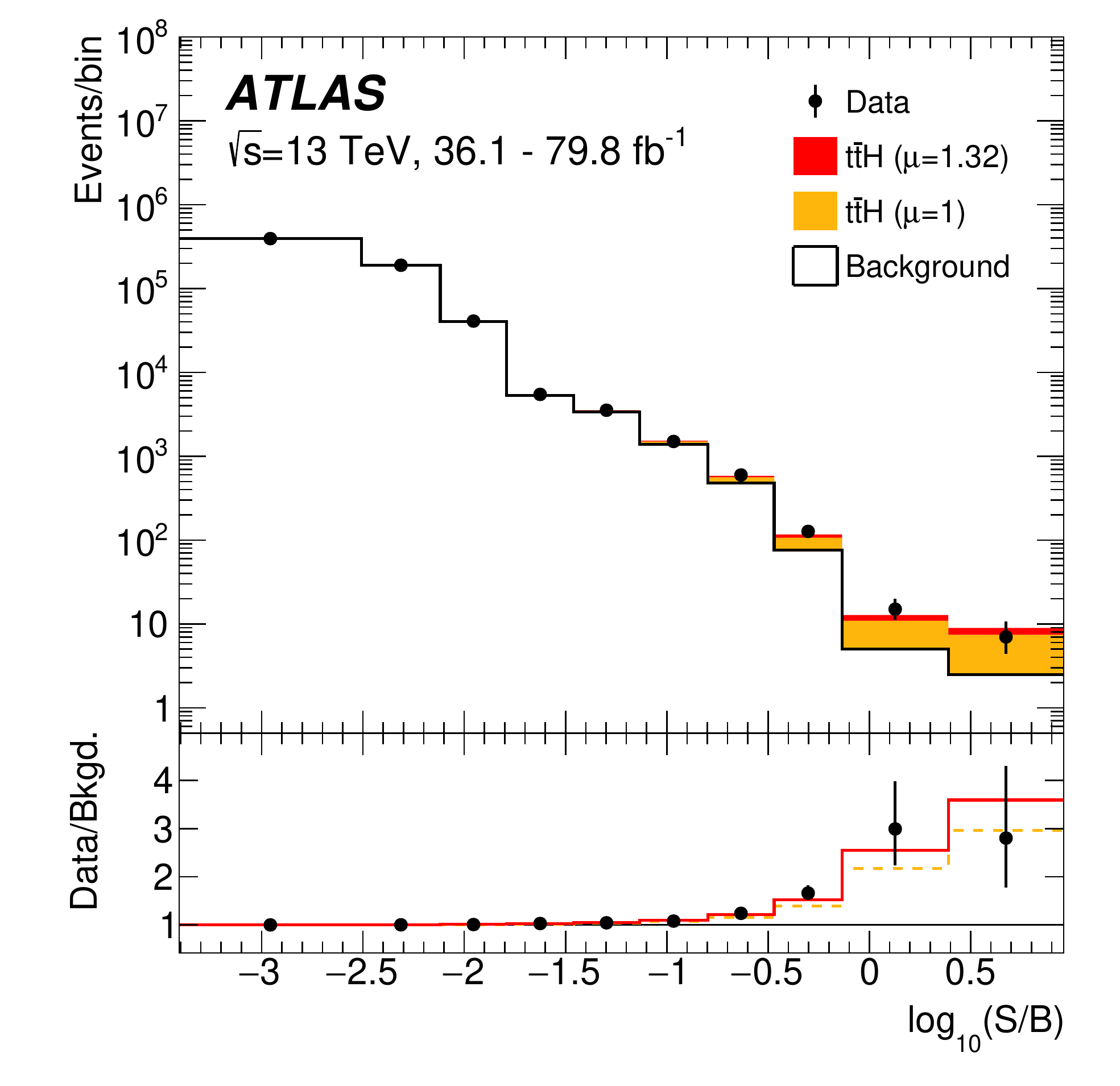}}
\subfigure[]{
\includegraphics[width=0.35\textwidth]{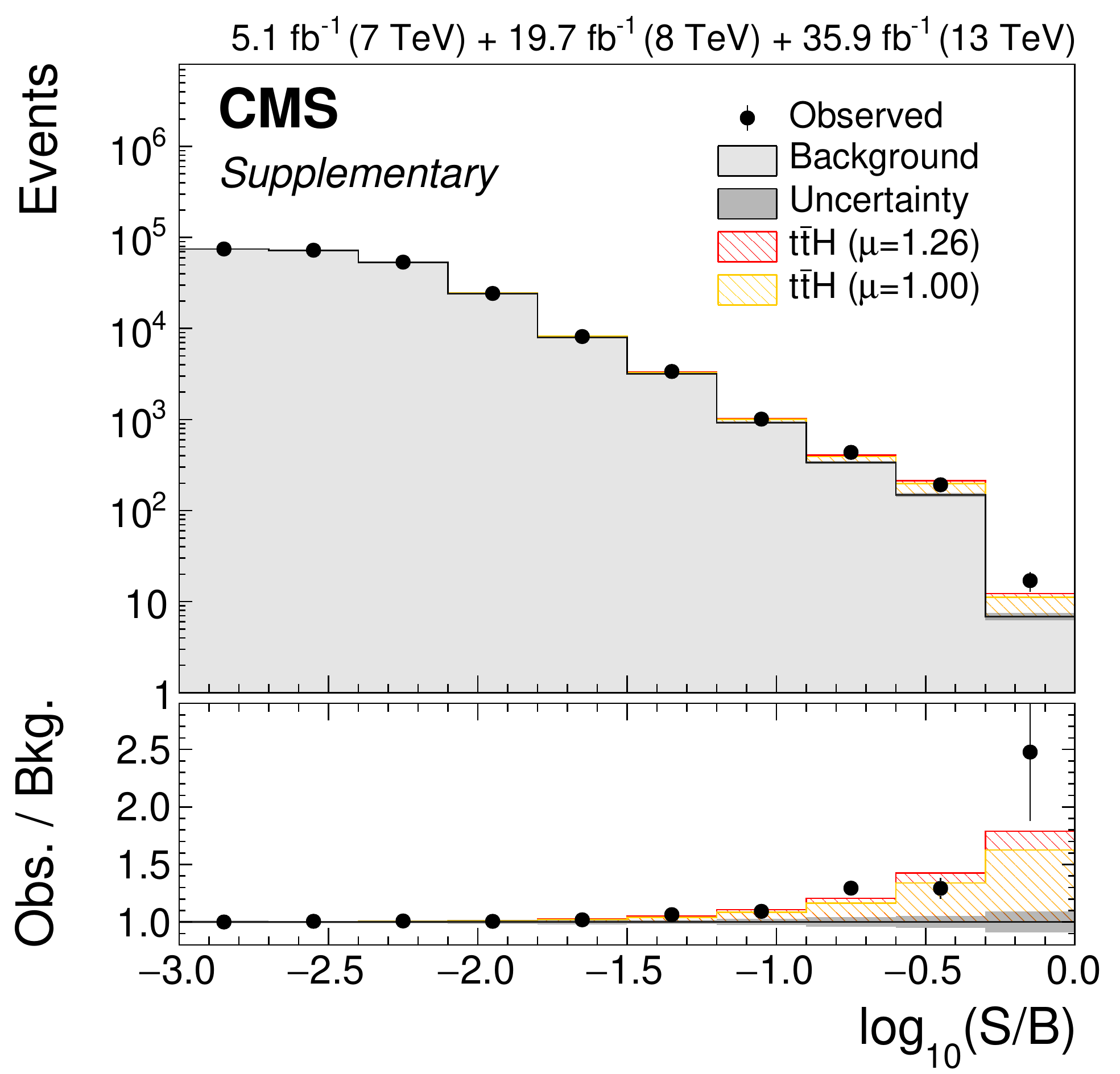}}
\caption{Distribution of all events in the $\ttbar H$~analyses as a function of $\log{\frac{s}{b}}$ for (a) the ATLAS~\cite{Aaboud:2018urx} and (b) CMS~\cite{Sirunyan:2018hoz} analyses.}
\label{f:ttH}
\end{figure}

The large top-quark mass in the SM is understood in terms of a large Yukawa coupling with the Higgs boson, which means 
the SM provides a prediction for the rate of the $\ttbar H$~process that is proportional to that coupling squared. The observation of this process,
which was obtained by both ATLAS~\cite{Aaboud:2018urx} and CMS~\cite{Sirunyan:2018hoz}, is therefore significant evidence of the fermion mass generation mechanism in the SM. The multiple decay modes
of the Higgs boson means that both collaborations have pursued a wide range of final states in order to maximise the sensitivity to the $\ttbar H$~process. Many
of these channels are also applying multi-variate techniques and a detailed description is beyond the scope of these proceedings.
The outcome of these complex searches is summarised in Figure~\ref{f:ttH}, which shows all the selected events ordered by $\log{\frac{s}{b}}$. The $ttH$~signal
is clearly visible in the tails of the distributions and this represents a milestone result that opens up detailed study of the top-Higgs coupling in the years ahead.

\section{Searching with top quarks}
\label{s:search}

One of the main goals of the LHC is to search for signs of new physics around the TeV scale. The large mass of the top-quark means that it is
often crucial in new physics models that seek to resolve the hierarchy problem. The huge array of searches is impressive, but with no sign yet
of significant deviations from the SM, we must be sure to cover as many corners as we can. At the workshop D. Shih motivated us to go
``harder, faster, better, stronger"~\cite{ShihDaftPunk} and I found it interesting to look for examples of such searches.

\begin{figure}[htb]
\centering
\subfigure[]{
\includegraphics[width=0.38\textwidth]{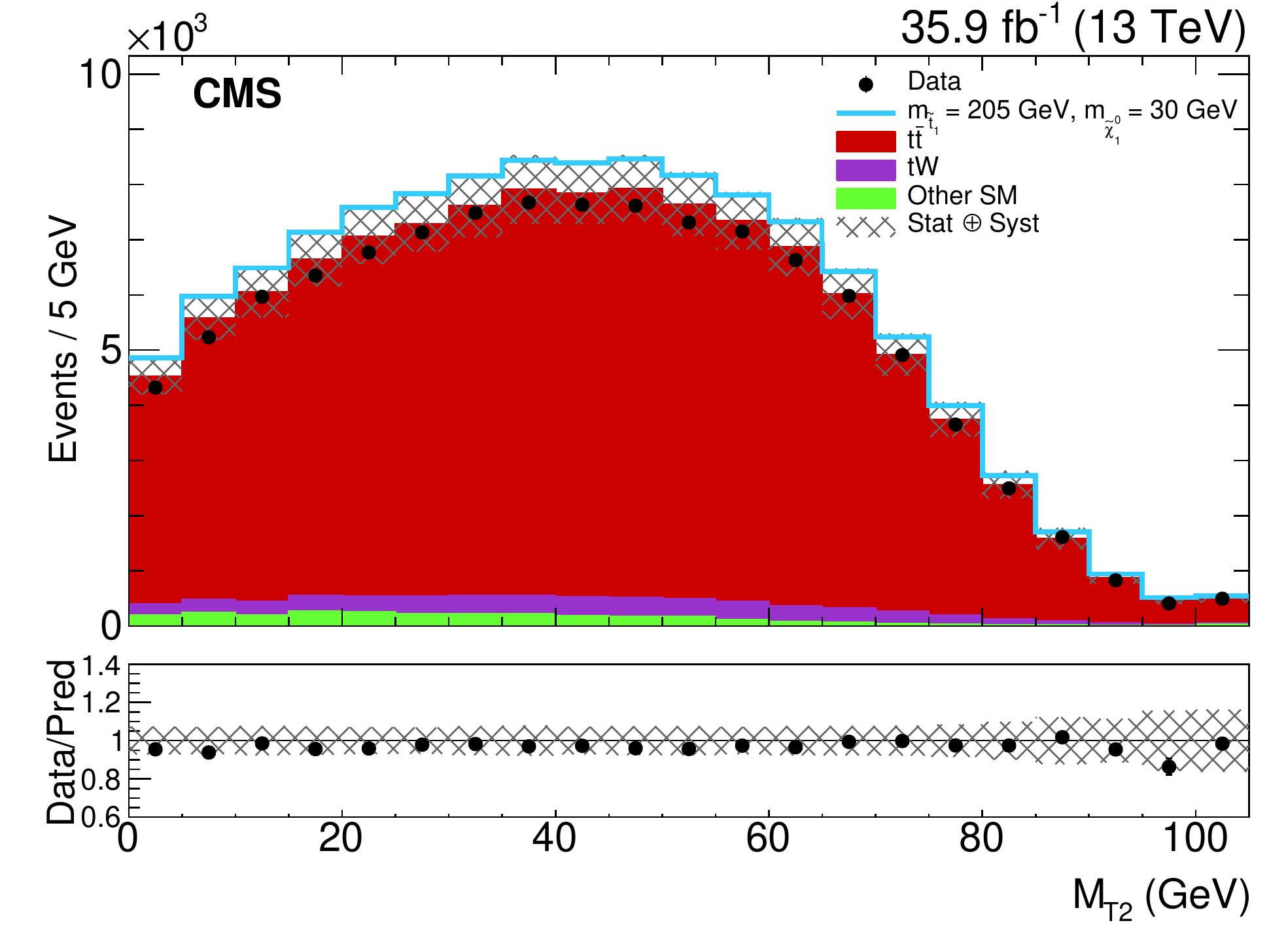}}
\subfigure[]{
\includegraphics[width=0.3\textwidth]{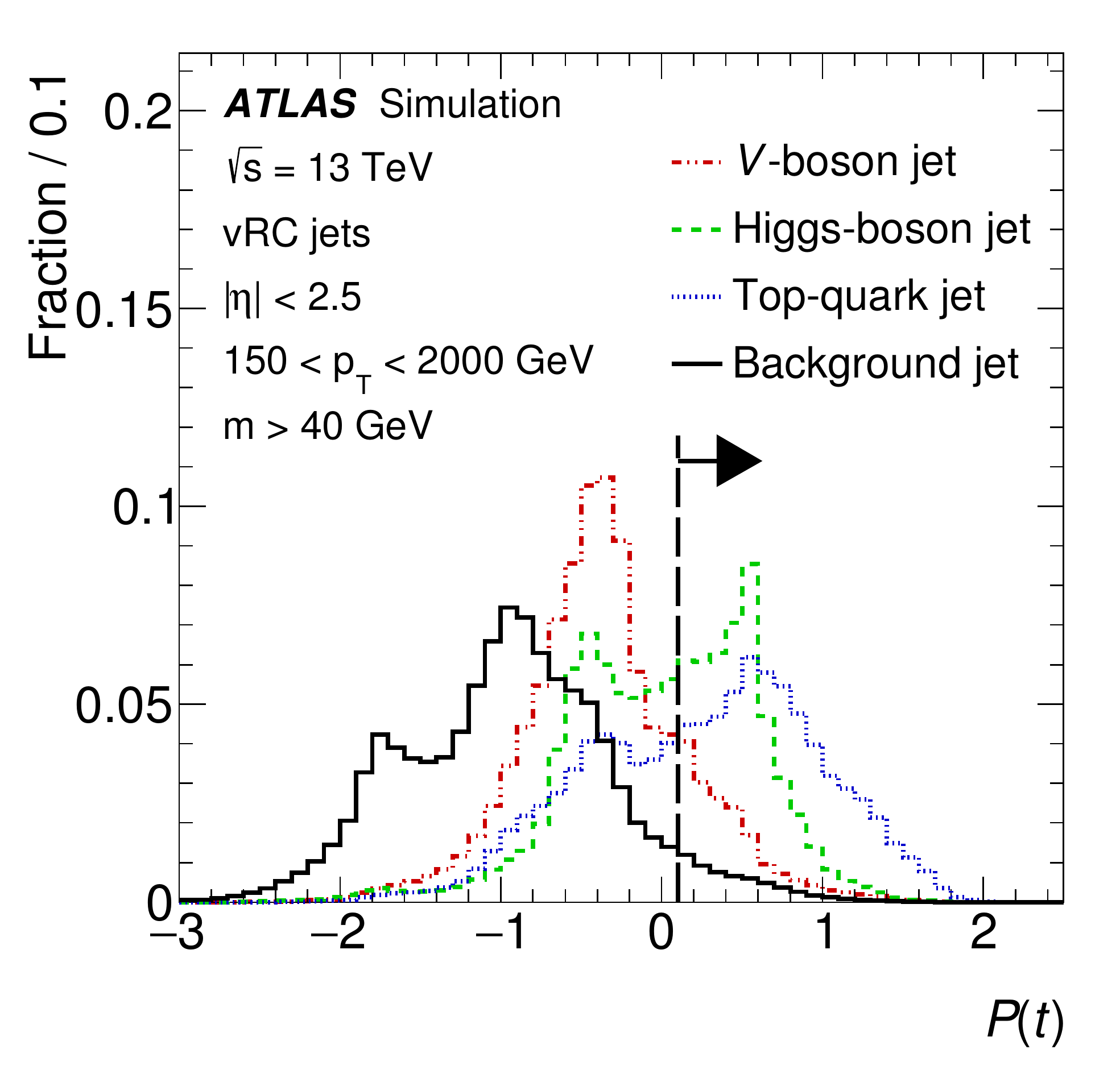}}
\caption{(a) $m_{T2}$~distribution used in the CMS search for light stop quarks~\cite{Sirunyan:2019zyu}.
(b) Deep neural network used to select highly boosted top-quarks in the ATLAS vector-like-quark search~\cite{Aaboud:2018wxv} .}
\label{f:stop}
\end{figure}

A nice example of
a ``harder" search is the recent CMS search for stop quarks that targets light stop quarks~\cite{Sirunyan:2019zyu}. This region is challenging, because with a light neutralino,
the final states are difficult to extract from the SM background. The CMS search exploits the $m_{T2}$~distribution, shown in Figure~\ref{f:stop}
to perform the search. The technique achieves good sensitivity, however we should be aware that we rely largely here on the theoretical modelling of the distribution,
which can be difficult, for example, due to off-shell effects that are not included in the current MC generators used by ATLAS and CMS. ``Faster" for an experimentalist
tends to be linked with producing results promptly after data has been collected. At the workshop, we saw a new search for charged lepton flavour violation
by the ATLAS experiment~\cite{ATLAS-CONF-2018-044}, that included the 2017 data. This also illustrates the ability of the experiments to react to new ideas, in this
case looking at lepton flavour violation with an eye on the anomalies seen in the flavour sector. Interpreting ``Better" is perhaps slightly subjective, but I decided to take
an example of an analysis using sophisticated techniques. The ATLAS search for vector-like-quarks~\cite{Aaboud:2018wxv} uses a deep neural network to select boosted
top, $W$~and $Z$~bosons (shown in Figure~\ref{f:stop}) and then uses a matrix-element-method to search for the signal above the SM background. The machine
learning techniques are likely to become more widely used in the future and hopefully will improve our sensitivity to new physics. Translating "Stronger" is perhaps
most tricky, but here I think it perhaps reminds us of the need to stay optimistic: just because we have not discovered new physics in the first phase of the LHC doesn't
mean there is not something fascinating to discover in the years ahead.

\section{The future and summary}
\label{s:future}

The workshop concluded with a forward look to future experiments. The high-luminosity LHC (HL-LHC) is the only approved future project and will offer an order of magnitude
improvement in statistics. The natural top-quark analyses that can benefit from this project are rare searches (e.g. FCNC), however a challenge in the years ahead will be to maximise the physics
output for the ``bulk" top-quark physics programme, including those areas that are systematics limited already in run-2.
Beyond the HL-LHC, several possibilities for precision $e^+e^-$~machines exist and offer extremely interesting top physics programmes, including a top mass
measurement below $100$~MeV~\cite{Abramowicz:2018rjq}. The success of all these machines will rely on the next-generation of physicists and with an eye on this,
I think it is great to see that the Top workshop continues to run both poster and young-scientist sessions.

Trying to summarise a summary always seems something that is hard to achieve. Perhaps the one overriding impression from the workshop is that we have a wealth of
data and the challenge ahead is to use that data to maximise our understanding of the fundamental physics.

%\clearpage

\Acknowledgements
I am very grateful to the organisers of the workshop for the invitation to given the experimental summary at the Top 2018 workshop
and for organising such a nice workshop. I'm also grateful for the funding from the Royal Society.
I would also like to thank all the speakers for their patience with all my questions.

\bibliographystyle{Science}
\bibliography{refs}

\end{document}